\documentstyle[12pt]{article}
\begin{document}

\begin{center}

{\Large \bf Current conservation in two-dimensional\\
\vspace{0.2in}
AC-transport}

\bigskip

Jian Wang and Qingrong Zheng

\bigskip

{\it
Department of Physics, \\
The University of Hong Kong,\\
Pokfulam Road, Hong Kong.
}

\bigskip

Hong Guo

\bigskip

{\it
Centre for the Physics of Materials,\\
Department of Physics, McGill University,\\
Montreal, Quebec, Canada H3A 2T8.
}

\end{center}

\vfill

\baselineskip 15pt               

The electric current conservation in a two-dimensional quantum 
wire under a time dependent field is investigated. 
Such a conservation is obtained as the global density of states
contribution to the emittance is balanced by the contribution due
to the internal charge response inside the sample. However when the
global partial density of states is approximately calculated 
using scattering matrix only, correction terms are needed to obtain
precise current conservation. We have derived these corrections
analytically using a specific two-dimensional system. We found that 
when the incident energy $E$ is near the first subband, our 
result reduces to the one-dimensional result. As $E$ approaches 
to the $n$-th subband with $n>1$, the correction term diverges. 
This explains the systematic deviation to precise current
conservation observed in a previous numerical calculation.

\vfill

\baselineskip 16pt

{PACS number: 72.10.Bg, 73.20.Dx, 73.40.Gk, 73.40.Lq}

\newpage
\section{Introduction}

The dynamic conductance of a quantum coherent mesoscopic system under 
a time dependent external field is the subject of recent 
interests\cite{but1,bruder,pieper,chen,wang1}. In contrast to 
dc-transport where the internal potential distribution inside the sample
does not appear explicitly, the AC-response depends sensitively on the internal
potential distribution. This internal potential is due to the charge 
distribution generated by the applied AC-field at the leads and it has
to be determined self-consistently\cite{but1}. So far there are
two approaches to the coherent AC-transport problem. One is to derive 
a formal linear response to a given potential distribution in the
sample\cite{baranger}. The difficulty with such an approach is that the 
potential distribution is not known a priori. Another approach is to 
investigate the AC-response to an external perturbation which prescribes 
the potentials in the reservoirs only\cite{pastawski,but1}. The
external potentials effectively determine the chemical potential of the
reservoirs and the potential distribution in the conductor must be
considered a part of the response which is to be calculated
self-consistently. In this approach, B\"uttiker and his
coworkers\cite{but1,but3} have formulated a current conserving 
formalism for the low frequency admittance of mesoscopic conductors. 

In the theory of B\"uttiker, Pr\^etre and Thomas\cite{but1}, it is 
necessary to consider the Coulomb interactions between the many 
charges inside the sample, in order to preserve the current conservation. 
For a multi-probe conductor the low frequency admittance is found to 
have the form\cite{but3,but4}
$G_{\alpha\beta}(\omega)=G_{\alpha\beta}(0)-i\omega E_{\alpha\beta}
+O(\omega^2)$, 
where $G_{\alpha \beta}(0)$ is the dc-conductance, $E_{\alpha \beta}$
is the emittance\cite{but3}, and $\alpha$ (or $\beta$) labels the probe. 
The emittance $E_{\alpha \beta}$ describes the current response at probe 
$\alpha$ due to a variation of the electro-chemical potential at probe 
$\beta$ to leading order with respect to frequency $\omega$. It can 
be written as\cite{but3,wang1}
$E_{\alpha \beta} = dN_{\alpha \beta}/dE - D_{\alpha \beta}$,
where the term $dN_{\alpha \beta}/dE$ is the global partial density of 
states (GPDOS)\cite{gas1} which is related to the scattering matrix. 
It describes the density of states of carriers injected in probe $\beta$ 
reaching probe $\alpha$ and is a purely kinetic term. The term 
$D_{\alpha \beta}$ is due to the Coulomb interaction of electrons 
inside the sample and is a term of capacitive nature. $D_{\alpha\beta}$ 
can be computed from the local density of states\cite{but1,but3} 
which is related to the electron dwell times. 
Electric current conservation, namely
$\sum_{\alpha}G_{\alpha\beta}(\omega)=0$, means that 
$\sum_{\alpha}E_{\alpha\beta}=0$ or equivalently\cite{but1,ian}
\begin{equation}
\frac{dN_{\beta}}{dE}\equiv \sum_{\alpha} \frac{dN_{\alpha \beta}}{dE} 
= \sum_{\alpha} D_{\alpha\beta}=\frac{\tau_{d,\beta}}{h}
\label{eq1}
\end{equation}
where $dN_{\beta}/dE$ is the DOS and $\tau_{d,\beta}$ is the dwell
time for particles coming from the probe $\beta$. Clearly the current
conservation is established since one realizes that 
$\sum_{\alpha}dN_{\alpha\beta}/dE$ is the physical quantity called
injectance which is identical\cite{but3} to 
$\sum_{\alpha}D_{\alpha\beta}$.

Applying the above formalism to mesoscopic conductors, one needs to 
compute various physical quantities\cite{wang1} such as the partial 
density of states. These quantities have vivid physical meaning\cite{wang1}
but are not easy to obtain exactly. For a large system, the GPDOS can 
be expressed approximately in terms of the energy derivative of the 
scattering matrix elements\cite{avishai}: 
\begin{equation}
\frac{dN_{\alpha\beta}}{dE} = \frac{1}{4\pi i} \left(
s_{\alpha\beta}^{\dagger}\frac{ds_{\alpha \beta}}{dE} - 
\frac{ds_{\alpha\beta}^{\dagger}}{dE}s_{\alpha\beta}\right)\ \ .
\label{eq2}
\end{equation}
Because for a given system one may be able to obtain the scattering matrix,
Eq.(\ref{eq2}) thus provides a practical means of computing the GPDOS.
On the other hand, in order to obtain current conservation 
{\it precisely}, a correction should be added to Eq. (\ref{eq2}) which 
can be neglected for large systems and large energies\cite{gas1,gas2}.
For one-dimensional systems, such a correction has been derived by
Gasparian {\it et. al.}\cite{gas2} which contains the reflection 
amplitude divided by the energy,
\begin{equation}
\frac{dN_{\alpha}}{dE} = \frac{d\bar{N}_{\alpha}}{dE} +
Im \{\frac{s_{\alpha \alpha}}{4\pi E} \}\ \ ,
\label{eq3}
\end{equation}
where $d\bar{N}_{\alpha}/dE\equiv \sum_{\beta}dN_{\alpha\beta}/dE$ which
is computed from Eq. (\ref{eq2}).

We have recently applied the above current conserving formalism 
to a {\it two-dimensional} mesoscopic conductor in the shape of a
T-junction\cite{wang1}. To the best of our knowledge, it was the first
2D calculation with first principles.  Among other things, an interesting
and we believe useful discovery was that Eq.(\ref{eq3}) turned out
to be inaccurate in 2D. First of all, energy $E$ in the second term on the
right hand side of Eq.(\ref{eq3}) has to be replaced by the longitudinal 
part of the incident energy.  Even with this change, there were small 
but systematic deviations to precise current conservation when the 
energy is approaching the second subband. In fact it was found that the 
DOS $d\bar{N}_{\alpha}/dE$ as defined above diverges near the onset of 
the second subband and this led to the observed systematic 
deviations\cite{wang1}.

We are not aware of any 2D theory to account for the correction term 
which should appear in Eq. (\ref{eq3}). The purpose of this paper is to
investigate such correction terms in two dimensions. 
This not only provides further theoretical insights to the problem of
AC-transport, but is also very helpful from a practical application 
point of view. From our own experience, numerical AC-transport calculations 
can be quite tricky and being able to obtain precise electric current 
conservation often serves as a very stringent check to numerical results.
To this purpose, we have considered the simplest two-dimensional model 
which is a $\delta$-potential inside a quasi-1D ballistic 
conductor\cite{bag}. Since quantum scattering in this system 
leads to mode mixing which is the basic feature of a two-dimensional 
system, it provides answers to our 2D problem. The advantage of this system
is that it can be solved exactly. We have thus derived analytically 
the correction term.  In particular we found that when the 
incident energy $E$ is within the first subband, our result essentially
reduces to the one-dimensional result Eq. (\ref{eq3}). As $E$ is increased
to approach the $n$-th subband edge with $n>1$, the correction term 
diverges. This explains the systematic deviation observed in our
previous numerical calculation\cite{wang1}.

The paper is organized as the following.  In the next section we present the
solution of the 2D scattering problem and derive the correction term. Section
III contains our numerical tests of the analytical formula. The
last section serves as a summary.

\section{Model and results}

Figure 1 shows the system where a $\delta$-potential is 
confined inside a quasi-1D wire with width $a$. We assume, for simplicity
of the calculation, that the boundaries of the ballistic conductor 
are hard walls, {\it i.e.} the potential $V=\infty$. Inside the conductor,
the potential is zero except that a $\delta$ function potential $V(x,y) =
\gamma \delta(x) \delta(y-y_0)$ is placed at $\vec{r}=(0,y_0)$. 
The scattering region $x_1 < x < x_2$ is assumed to be
symmetric with $x_2 = -x_1 =L/2$. From now on we set 
$\hbar= 1$ and $m=1/2$ to fix our units.

To compute the transmission and reflection amplitudes thus the scattering 
matrix, a mode matching method\cite{schult} is employed. 
The electron wave functions are written as follows. For region I 
(see figure 1): 
\[\Psi_I = \sum_n \chi_n(y) \left(a_n e^{i k_n x} + b_n e^{-i k_n x} 
\right)    \ \ , \]
where $\chi_n(y)$ is the wave function of the $n$-th subband 
along y-direction; $a_n$ is the incoming wave amplitude and taken as 
an input parameter; $b_n$ is the reflection amplitude;
and $k_n^2 = E - (n\pi/a)^2$ is the longitudinal momentum for the $n$-th 
mode. Note that for electron traveling in the first subband, 
$k_n$ with $n>1$ is purely imaginary. Similarly for region II:
\[\Psi_{II} = \sum_n \chi_n(y) \left(c_n e^{i k_n x} + d_n e^{-i k_n x} 
\right) \ \ , \]
where $c_n$ is transmission amplitude and $d_n$ is set to zero in our 
calculation. After matching the boundary conditions at $x=0$, we obtain
\[a_n+b_n = c_n\]
and
\[i k_n c_n - i k_n (a_n - b_n) = \sum_m \Gamma_{nm} (a_m +b_m) \ \ ,\]
where $\Gamma_{nm} = \gamma \chi^*_n(y_0) \chi_m(y_0)$. Eliminating 
$c_n$, we have
\begin{equation}
\vec{e} = P \vec{b} \ \ ,
\end{equation}
where $e_n = -\sum_m \Gamma_{nm} a_m$ and $P_{nm} = \Gamma_{nm} -2 i k_n
\delta_{nm}$. To find $\vec{b}$ we need to compute $P^{-1}$.
Introducing a new matrix $\tilde{P} \equiv I + M$ 
with $M_{nm} = i \Gamma_{nm}/(2k_m)$ so that $\tilde{P}_{nm}
(-2ik_m) = P_{nm}$. Expanding $\tilde{P}^{-1}$ in powers of $M$, we have
\[\tilde{P}^{-1} = \frac{1}{I+M} = I - M + M^2 - M^3 ...\]
Since $\Gamma_{nm} \Gamma_{ml} = \Gamma_{nl} \Gamma_{mm}$, we find that
$M^2 = (\alpha -1) M $ where $\alpha = 1+i \sum_n \Gamma_{nn}/(2k_n)$,
from which we have $\tilde{P}^{-1} = 1-M/\alpha$. Finally, we obtain 
the matrix elements
\begin{equation}
(P^{-1})_{nm} = \frac{i}{2k_n} (\delta_{nm} - \frac{i \Gamma_{nm}}
{2 k_m \alpha} )\ \ .
\label{eqp}
\end{equation}

We shall specialize to consider the incident electron being in the 
first subband: $a_n = \delta_{n1}$.  Using Eq.(\ref{eqp}) the reflection 
and transmission amplitudes are
\begin{equation}
b_n = \sum_m (P^{-1})_{nm} e_m = \frac{-i \Gamma_{n1}}{2k_n \alpha}\ \ ,
\end{equation}
\begin{equation}
c_n = \delta_{n1} +b_n\ \ .
\end{equation}
For our system the scattering matrix elements $s_{\alpha \beta}$ are 
given by $s_{11} = b_1 \exp(ik_1 L)$ and $s_{12} = c_1 \exp(i k_1 L)$. 
The approximate DOS becomes, using Eq. (\ref{eq2}),
\begin{eqnarray}
\frac{d\bar{N}_{\alpha}}{dE} & = &
\frac{1}{4\pi i} \sum_{\beta}\left(s_{\alpha
\beta}^{\dagger} \frac{ds_{\alpha \beta}}{dE} - 
\frac{ds_{\alpha\beta}^{\dagger}}{dE} s_{\alpha\beta} \right)
\nonumber \\
& = &\frac{L}{4\pi k_1} - Im\left(\frac{b_1}{4\pi k_1^2}\right) - 
\frac{1}{4\pi} \sum_n \frac{|b_n|^2}{i k_1 k_n}  \ \ .
\label{dosbad}
\end{eqnarray}
To derive this expression we have used a relation $2b_1^* +1 = \alpha 
/\alpha^*$ which follows directly from the unitary condition of the 
scattering matrix. Next we compute the dwell time and hence the 
precise DOS (as opposed to the approximate DOS of Eq. (\ref{dosbad})):
\begin{eqnarray}
\tau_{d,1} &=& \int_I |\Psi_I|^2 dx dy + \int_{II} |\Psi_{II}|^2 dx dy 
\nonumber \\
& = & \frac{L}{2k_1} + Re (b_1 \frac{e^{i k_1 L}-1}{2ik_1^2})
+ \sum_n |b_n|^2 \frac{e^{ik_n L}-1}{2ik_1 k_n}\ \ .
\label{dosgood}
\end{eqnarray}

From Eqs.(\ref{eq1}),(\ref{dosbad}) and (\ref{dosgood}), we arrive at the
following central result of this work,
\begin{equation}
\frac{dN_{\alpha}}{dE} = \frac{d\bar{N}_{\alpha}}{dE} + 
Im \{\frac{s_{\alpha \alpha}}{4\pi k_1^2}\} +
\frac{1}{4\pi} \sum_{n=2} \frac{|b_n|^2}{i k_1 k_n} e^{i k_n L}\ \ .
\label{result}
\end{equation}
Hence we found that for this 2D system, there are two correction terms to
the DOS. Clearly the first correction term, {\it i.e.} the 2nd term on 
the right hand side of Eq. (\ref{result}), is generic, as it can be written 
in terms of the scattering matrix element. This term is similar
to the corresponding term in Eq. (\ref{eq3}) of the 1D case, 
except that the total energy $E$ in Eq. (\ref{eq3}) is now replaced by the 
transport energy $k_1^2$. In fact this them has been guessed in our earlier
work\cite{wang1}. There is a second correction term (the 3rd term of
Eq.(\ref{result})) which comes solely due to mode mixing in our 2D system,
and understandably it does not exist in 1D cases\cite{gas2}.

For small incident energies, {\it i.e.} as $k_1$ goes to zero, 
$|b_n|^2 \rightarrow k_1^2$ for $n>1$. Therefore the second correction 
term of (\ref{result}) is actually negligible at small energies. Indeed,
this is the case in our earlier numerical calculations\cite{wang1} where
current conservation was very well satisfied at low energies using
Eq.(\ref{eq3}). However, as energy is approaching 
the $n$-th subband edge, for small $k_n \rightarrow 0$ with $n>1$, 
$|b_n|^2$ remains finite. Hence according to Eq. (\ref{result}) 
the second correction term diverges at these higher subband edges. 
This explains the observation of our calculation\cite{wang1} where
systematic numerical errors exist in current conservation near the 2nd
subband edge. For energies within the first subband, as
mentioned above $k_n$ are all pure imaginary numbers with $n>1$.
Hence with large system size $L$, the factor $\exp(i k_n L)$
is very small as long as $k_n\neq 0$. However we emphasis that 
the second correction term becomes dominant very near each subband edge 
thus must be included in order to obtain precise current conservation.

Finally we note that Eq.(\ref{result}) is not coordinate independent, 
so care must be taken when using it. For instance, if we choose 
$x_1$ as the origin in figure 1, the factor
$\exp(i k_n L)$ in the last term of Eq.(\ref{result}) will be 
canceled due to the coordinate shift while the second term of 
Eq.(\ref{result}) remains the same. In this sense, the new correction term
is not generic and must be computed case by case for 2D systems.

\section{Numerical test}

To gain further intuitive impression of the AC-transport, and in particular
to check our analytical formula Eq.(\ref{result}), we shall
first present direct numerical calculations of the
admittance for the quantum wire system studied in the last
section (Fig.(1)).  Obviously since this problem was solved
exactly above, agreement is obtained with Eq.(\ref{result}).  
We shall then study the validity of Eq.(\ref{result}) using another 
more complicated 2D conductor in the shape of a T-junction (see below). 
Indeed, although Eq.(\ref{result}) was derived using a specific example of
Fig.(1), it dramatically improves the current conservation near
the second subband edge for the T-junction as well.

In order to compute the admittance, we have to know $D_{\alpha, \beta}$
which is related to the dwell time\cite{but1,wang1}. For a metallic
conductor, it is appropriate to use the Thomas-Fermi approximation.
Under such an approximation $D_{\alpha, \beta}$ is given by\cite{but1,but3}
\begin{equation}
D_{\alpha, \beta} = \int d^3r \frac{(dn(\alpha, \vec{r})/dE )
 (dn(\vec{r}, \beta)/dE)} {dn(\vec{r})/dE} \ \ ,
\label{d11}
\end{equation}
where the local density of states $dn(\vec{r},\beta)/dE$ is the injectivity 
which measures the additional local charge density brought into the sample 
at point $\vec{r}$ by the oscillating chemical potential at probe $\beta$. 
The injectivity can be expressed as\cite{but1}
\begin{equation}
\frac{dn(\vec{r}, \beta)}{dE} = \sum_n \frac{|\Psi_{\beta
n}(\vec{r})|^2} {2\pi v_{\beta n}} \ \ ,
\label{dn}
\end{equation}
where $v_{\beta n}$ is the velocity of carriers at the Fermi energy at
mode $n$ in probe $\beta$. $dn(\alpha, \vec{r})/dE$ is called the emissivity 
which describes the local density of states of carriers at point $\vec{r}$
which are emitted by the conductor at probe $\alpha$. It is defined as
\begin{equation}
\frac{dn(\alpha, \vec{r})}{dE} = -\frac{1}{4\pi i} \sum_{\beta} Tr \left[
s_{\alpha \beta}^{\dagger} 
\frac{\delta s_{\alpha \beta}}{e \delta U(\vec{r})}  - 
\frac{\delta s_{\alpha \beta}^{\dagger}}{e \delta U(\vec{r})} 
s_{\alpha \beta} \right] \ \ .
\end{equation}
It has been shown\cite{but4} that in the absence of magnetic field the
injectivity is equal to the emissivity. Using Eqs.(\ref{dosbad}), 
(\ref{d11}) and (\ref{dn}), we can calculate the emittance.

Specifically, for the system of Fig.(1) we consider incident 
electron coming from probe 1 and set $a=L=1$, $y_0=0.3$, and $\gamma=-1$. 
In Fig.(2), we plot the global
DOS together with the transmission coefficient $T$. As expected, the
transmission coefficient $T(E)$ (solid line) has large values for almost all 
energies $E$ except at a special energy $E_r$ where we have complete reflection 
(reflection coefficient $R(E_r)=1$) due to the resonant state. This can also be 
seen from the behavior of the global partial DOS for reflection $dN_{11}/dE$ 
(dotted line) which peaks when $T(E=E_r)=0$. On the other hand, $dN_{21}/dE$ 
(dashed line), which is the global partial DOS for transmission, takes 
minimum value at $E=E_r$. This behavior is consistent with that of a 1D 
system made of a symmetric scatterer\cite{gas1} where one has $dN_{11}/dE 
\sim R dN/dE$ and $dN_{21}/dE\sim T dN/dE$. In Fig.(3), the
quantities $D_{11}$ (solid line) and $D_{12}$ (dotted line) are shown. 
Both curves reach maximum values near the resonant point $E_r$, which is
expected since $D_{\alpha \beta}$ are proportional to the dwell time or
the DOS. The emittance $E_{\alpha \beta}$ is plotted in Fig.(4).
Both $E_{11}$ (solid line) and $E_{12}$ (dotted line) reach extremal 
values at the resonant point. The system responds differently for 
different energy, either capacitively when $E_{11}=-E_{12} >0$, 
or inductively otherwise. From Fig.(4), we observe that near 
the resonance $E_{11}$ and $E_{12}$ respond capacitively while $E_{12}$ 
is inductive away from this resonance energy. This behavior, namely being 
capacitive when at the $T\approx 0$ resonance, is the same as that
observed in the 2D T-junction\cite{wang1}. On the other hand for 
an 1D tunneling system\cite{but1} the response is inductive at its resonance.
But in that case the resonance is marked by transmission coefficient being
near unity. Finally, to confirm electric current conservation, essentially
the two curves of Fig.(4) must add to zero. Clearly these curves
do not cancel each other as the figure shows, exactly due to the approximate
nature of the partial density of states as obtained using 
Eq. (\ref{eq2}). After including the two corrections to DOS
as derived in Eq. (\ref{result}), however, we did obtain perfect current
conservation for the whole energy range.  This is not surprising since after
all (\ref{result}) is an exact result for this quantum system.

Our main result Eq.(\ref{result}) is derived using a specific simple
example shown in Fig.(1). There seems no special reason for 
Eq.(\ref{result}) to apply to other 2D systems, since the form of the
new correction term is given by the amplitudes of the non-propagating 
modes inside the scattering junction (as oppose to the more general 
scattering matrix elements), and these evanescent amplitudes probably 
depend on the scatterer in some fashion. In this sense it is 
unfortunate that a more general form was not obtained. However since 
the new correction term does explain, qualitatively, the 
observed\cite{wang1} discrepancy of using Eq. (\ref{eq3}) as 
discussed above, it is tempting to test it using the more complicated 
2D system of the T-junction studied previously\cite{wang1}. As the 
T-junction has been reviewed and studied by many authors\cite{sols,wang1} 
at various contexts, here we shall not present the details for its 
calculation.  For this purpose, we have checked the current
conservation of the T-junction using Eq.(\ref{eq3}) and compared with 
the result obtained using Eq.(\ref{result}). In Fig.(5), 
we have plotted the DOS $d\bar{N}_1/dE$ given by Eq.(\ref{eq3}) 
(dotted line) and by\cite{foot1} Eq. (\ref{result}) (solid line), and the 
dwell time $\tau_{d,1}/2\pi$ (dashed line). Although the result 
Eq. (\ref{result}) is model dependent, we observe that the agreement is 
clearly better. This suggests that the new correction term does capture 
the essential ingredient of the correction, although it is not completely 
universal as the evanescent amplitudes depend on the peculiarities of a 2D 
system in some weak way, leading to the small remaining difference.

\section{Summary}

In summary, we have investigated the electric current conservation 
in a two-dimensional ballistic conductor under a time dependent 
field.  Similar to that of the 1D case, we found that in order to 
obtain precise current conservation, certain corrections to the
density of states as obtained approximately from the scattering matrix
must be included. We have derived these corrections analytically 
for a specific two-dimensional system and found that
there are two correction terms. One of the correction term has the
same form as that of the 1D case, while the second correction term
is purely due to mode mixing characteristic of 2D quantum scattering.
In particular, when the incident energy $E$ is within the first subband, 
our result essentially reduces to the one-dimensional result 
if $E$ is not too high. On the other hand as $E$ approaches to the 
$n$-th subband with $n>1$, the correction term diverges at the subband
edges. Hence in 2D the mode mixing leads to important changes in the 
global density of states and must be included if precise electric
current conservation is desired.  Finally, the new correction term
found here provides a qualitative explanation for the small but
systematic deviation to precise current conservation observed in our
previous numerical calculations\cite{wang1} on a 2D quantum wire in the shape
of the T-junction. Indeed, our numerical test has produced better 
agreement when the new formula derived here is used.

\section*{Acknowledgments}

We thank Prof. M. B\"uttiker for helpful communications and discussions.
We gratefully acknowledge support by a RGC grant from the Government of 
Hong Kong under grant number HKU 261/95P, a research grant from the 
Croucher Foundation, the Natural Sciences and Engineering Research 
Council of Canada and le Fonds pour la Formation de Chercheurs 
et l'Aide \`a la Recherche de la Province du Qu\'ebec.
We thank the Computer Center of the University of Hong Kong for
computational facilities.

\section*{Figure Captions}

\begin{itemize}

\item[{Figure 1.}] Schematic plot of the quantum wire system: a 
$\delta$ potential $\gamma\delta(\vec{r}-\vec{r_0})$ is confined 
inside a quasi-1D quantum wire, with $\vec{r_0} = (0,y_0)$. 
The wire width is $a$. The scattering region is between $x_1$ and $x_2$,
where $x_2=-x_1=L/2$. In our calculations, the parameters are
set to $L=a=1$, $y_0=0.3$, and $\gamma=-1.0$.

\item[{Figure 2.}] The global partial density of states and the transmission
coefficient as functions of electron energy $E$. Solid line: transmission 
coefficient $T$;  dotted line: $dN_{11}/dE$;  dashed line: $dN_{21}/dE$.
Unit of energy is $\hbar^2/2ma^2$.

\item[{Figure 3.}] The current response to the internal potential,
$D_{\alpha\beta}$, as a function of energy $E$.  Solid line: $D_{11}$; 
dotted line: $D_{21}$.  

\item[{Figure 4.}] The dynamic part of the admittance,
$E_{\alpha\beta}\equiv dN_{\alpha\beta}/dE-D_{\alpha\beta}$ 
as a function of energy.

\item[{Figure 5.}] A numerical check of the electric current conservation,
Eq. (\ref{eq1}), for the T-junction studied in Ref. \cite{wang1}.
Solid line: $dN_1/dE$ as obtained by Eq. (\ref{result}); 
dotted line: $\tau_d/h\ =\ \sum_{\alpha}D_{\alpha 1}/h$.
Agreement of the two curves indicate the conservation.
The remaining small differences at high end of the energy between
the two curves indicates that the new correction term in
Eq. (\ref{result}) has a weak non-universal dependence on the 2D 
system shapes.

\end{itemize}

\begin{thebibliography}{00}
\bibitem{foot1}
To apply Eq. (\ref{result}) for the T-junction, we have substituted 
the appropriate amplitudes $b_n$ for the T-junction into (\ref{result}).
\end{thebibliography}
\end{document}